\documentclass[aps]{article}
\usepackage[T1]{fontenc}
\usepackage[latin1]{inputenc}
\usepackage{graphics}
\usepackage{subfigure}

\makeatletter

\providecommand{\LyX}{L\kern-.1667em\lower.25em\hbox{Y}\kern-.125emX\@}

\usepackage[T1]{fontenc}
\usepackage[latin1]{inputenc}


\begin{document}

\title{Inequalities for dealing with detector inefficiencies in Greenberger-Horne-Zeilinger-type
experiments\thanks{
Copyright (2000) by the American Physical Society. To appear in \emph{Phys.
Rev. Lett.}
}}

\author{\textbf{J. Acacio de Barros}\thanks{
E-mail: barros@ockham.stanford.edu. On leave from Dept. de F\'{\i}sica -- ICE,
UFJF, 36036-330 MG, Brazil. 
} and \textbf{Patrick Suppes}\thanks{
E-mail: suppes@ockham.stanford.edu. 
}\\
 CSLI - Ventura Hall \\
Stanford University \\
Stanford, CA 94305-4115}

\maketitle
\begin{abstract}
In this article we show that the three-particle GHZ theorem can be reformulated
in terms of inequalities, allowing imperfect correlations due to detector inefficencies.
We show quantitatively that taking into account these inefficiencies, the published
results of the Innsbruck experiment support the nonexistence of local hidden
variables that explain the experimental results. 
\end{abstract}
The issue of the completeness of quantum mechanics has been a subject of intense
research for almost a century. Recently, Greenberger, Horne and Zeilinger (GHZ)
proposed a new test for quantum mechanics based on correlations between more
than two particles \cite{GHZ}. What makes the GHZ proposal distinct from Bell's
inequalities is that they use perfect correlations that result in mathematical
contradictions. The argument, as stated by Mermin in \cite{Mermin}, goes as
follows. We start with a three-particle entangled state 
\[
|\psi \rangle =\frac{1}{\sqrt{2}}(|+\rangle _{1}|+\rangle _{2}|-\rangle _{3}+|-\rangle _{1}|-\rangle _{2}|+\rangle _{3}).\]
 This state is an eigenstate of the following spin operators: 
\begin{eqnarray}
\hat{\mathbf{A}} & = & \hat{\sigma }_{1x}\hat{\sigma }_{2y}\hat{\sigma }_{3y},\, \, \, \, \hat{\mathbf{B}}=\hat{\sigma }_{1y}\hat{\sigma }_{2x}\hat{\sigma }_{3y},\nonumber \\
\hat{\mathbf{C}} & = & \hat{\sigma }_{1y}\hat{\sigma }_{2y}\hat{\sigma }_{3x},\, \, \, \, \hat{\mathbf{D}}=\hat{\sigma }_{1x}\hat{\sigma }_{2x}\hat{\sigma }_{3x}.\nonumber 
\end{eqnarray}
 From the above we have that the expected correlations \( E(\hat{\mathbf{A}})=E(\hat{\mathbf{B}})=E(\hat{\mathbf{C}})=1. \)
However, \( \hat{\mathbf{D}}=\hat{\mathbf{A}}\hat{\mathbf{B}}\hat{\mathbf{C}}, \)
and we also obtain that, according to quantum mechanics, \( E(\hat{\mathbf{D}})=E(\hat{\mathbf{A}}\hat{\mathbf{B}}\hat{\mathbf{C}})=-1. \)
It is easy to show that these correlations yield a contradiction if we assume
that spin exist independent of the measurement process.

GHZ's proposed experiment, however, has a major problem. How can one verify
experimentally predictions based on perfect correlations? This was also a problem
in Bell's original paper. To ``avoid Bell's experimentally unrealistic restrictions'',
Clauser, Horne, Shimony and Holt \cite{Clauseretal} derived a new set of inequalities
that would take into account imperfections in the measurement process. A main
purpose of this article is to derive a set of inequalities for the experimentally
realizable GHZ correlations. We show that the following four inequalities are
both necessary and sufficient for the existence of a local hidden variable,
or, equivalently \cite{suppeszannoti}, a joint probability distribution of
\( \mathbf{A} \), \( \mathbf{B} \), \( \mathbf{C} \), and \( \mathbf{ABC} \),
where \( \mathbf{A},\mathbf{B},\mathbf{C} \) are three \( \pm 1 \) random
variables. 
\begin{equation}
\label{inequality1}
-2\leq E(\mathbf{A})+E(\mathbf{B})+E(\mathbf{C})-E(\mathbf{ABC})\leq 2,
\end{equation}
\begin{equation}
-2\leq -E(\mathbf{A})+E(\mathbf{B})+E(\mathbf{C})+E(\mathbf{ABC})\leq 2,
\end{equation}
\begin{equation}
-2\leq E(\mathbf{A})-E(\mathbf{B})+E(\mathbf{C})+E(\mathbf{ABC})\leq 2,
\end{equation}
\begin{equation}
\label{inequality4}
-2\leq E(\mathbf{A})+E(\mathbf{B})-E(\mathbf{C})+E(\mathbf{ABC})\leq 2.
\end{equation}

For the necessity argument we assume there is a joint probability distribution
consisting of the eight atoms \( abc,\ldots ,\overline{a}\overline{b}\overline{c} \),
where we use a notation where \( a \) is \( \mathbf{A}=1 \), \( \overline{a} \)
is \( \mathbf{A}=-1 \), and so on. Then, \( E(\mathbf{A})=P(a)-P(\overline{a}) \),
where \( P(a)=P(abc)+P(a\overline{b}c)+P(ab\overline{c})+P(a\overline{b}\overline{c}) \),
and \( P(\overline{a})=P(\overline{a}bc)+P(\overline{a}\overline{b}c)+P(\overline{a}b\overline{c})+P(\overline{a}\overline{b}\overline{c}) \),
and similar equations hold for \( E(\mathbf{B}) \) and \( E(\mathbf{C}) \).
Next we do a similar analysis of \( E(\mathbf{ABC}) \) in terms of the eight
atoms. Corresponding to (\ref{inequality1}), we now sum over the probability
expressions for the expectations \( F=E(\mathbf{A})+E(\mathbf{B})+E(\mathbf{C})-E(\mathbf{ABC}) \),
and obtain 
\begin{eqnarray}
F & = & 2[P(abc)+P(\overline{a}bc)+P(a\overline{b}c)+P(ab\overline{c})]\nonumber \\
 &  & -2[P(\overline{a}\overline{b}\overline{c})+P(\overline{a}\overline{b}c)+P(\overline{a}b\overline{c})+P(a\overline{b}\overline{c})].\nonumber 
\end{eqnarray}
 Since all the probabilities are nonnegative and sum to \( \leq 1 \), we infer
(\ref{inequality1}) at once. The derivation of the other three inequalities
is similar. To prove the converse, i.e., that these inequalities imply the existence
of a joint probability distribution, is slightly more complicated. We restrict
ourselves to the symmetric case \( P(a)=P(b)=P(c)\equiv p \), \( P(\mathbf{ABC}=1)\equiv q \)
and thus \( E(\mathbf{A})=E(\mathbf{B})=E(\mathbf{C})=2p-1, \) \( E(\mathbf{ABC})=2q-1. \)
In this case, (\ref{inequality1}) can be written as \( 0\leq 3p-q\leq 2, \)
while the other three inequalities yield just \( 0\leq p+q\leq 2 \). Let \( x\equiv P(\overline{a}bc)=P(a\overline{b}c)=P(ab\overline{c}) \),
\( y\equiv P(\overline{a}\overline{b}c)=P(\overline{a}b\overline{c})=P(a\overline{b}\overline{c}) \),
\( z\equiv P(abc) \) and \( w\equiv P(\overline{a}\overline{b}\overline{c}) \).
It is easy to show that on the boundary \( 3p=q \) defined by the inequalities
the values \( x=0,y=\frac{q}{3},z=0,w=1-q \) define a possible joint probability
distribution, since \( 3x+3y+z+w=1 \). On the other boundary, \( 3p=q+2 \)
a possible joint distribution is \( x=\frac{(1-q)}{3},y=0,z=q,w=0 \). Then,
for any values of \( q \) and \( p \) within the boundaries of the inequality
we can take a linear combination of these distributions with weights \( \frac{3p-q}{2} \)
and \( 1-\frac{3p-q}{2} \) and obtain the joint probability distribution, \( x=(1-\frac{3p-q}{2})\frac{1-q}{3},y=\frac{3p-q}{2}\frac{q}{3},z=(1-\frac{3p-q}{2})q,w=\frac{3p-q}{2}(1-q) \),
which proves that if the inequalities are satisfied a joint probability distribution
exists, and therefore a local hidden variable as well. The generalization to
the asymmetric case is tedious but straightforward.

The correlations present in the GHZ state are so strong that even if we allow
for experimental errors, the non-existence of a joint distribution can still
be verified. Let

\begin{description}
\item [(i)]\( E(\mathbf{A})=E(\mathbf{B})=E(\mathbf{C})\geq 1-\epsilon  \), 
\item [(ii)]\( E(\mathbf{ABC})\leq -1+\epsilon  \), 
\end{description}
where \( \epsilon  \) represents a decrease of the observed correlations due
to experimental errors. To see this, let us compute the value of \( F \) defined
above, \( F=3(1-\epsilon )-(-1+\epsilon ). \) But the observed correlations
are only compatible with a local hidden variable theory if \( F\leq 2 \), hence
\( \epsilon <\frac{1}{2}. \) Then, in the symmetric case, there cannot exist
a joint probability distribution of \( \mathbf{A},\, \mathbf{B} \) and \( \mathbf{C} \)
satisfying (i) and (ii) if \( \epsilon <1/2. \)

We will give an analysis of what happens to the correlations when the detectors
have efficiency \( d\in \lbrack 0,1] \) and a probability \( \gamma  \) of
detecting a dark photon within the window of observation when no real photon
is detected. Our analysis will be based on the experiment of Bouwmeester et
al. \cite{Innsbruck}. In their experiment, an ultraviolet pulse hits a nonlinear
crystal, and pairs of correlated photons are created. There is also a small
probability that two pairs are created within a window of observation, making
them indistinguishable. When this happens, by restricting to states where only
one photon is found on each output channel to the detectors, we obtain the following
state, 
\[
\frac{1}{\sqrt{2}}|+\rangle _{T}(|+\rangle _{1}|+\rangle _{2}|-\rangle _{3}+|-\rangle _{1}|-\rangle _{2}|+\rangle _{3}),\]
 where the subscripts refer to the detectors and \( + \) and \( - \) to the
linear polarization of the photon. Hence, if a photon is detected at the trigger
\( T \) (located after a polarizing beam splitter) the three-photon state at
detectors \( D_{1},D_{2} \), and \( D_{3} \) is a GHZ-correlated state (see
FIG. 1).
\begin{figure}
{\par\centering \resizebox*{3in}{!}{\rotatebox{-90}{\includegraphics{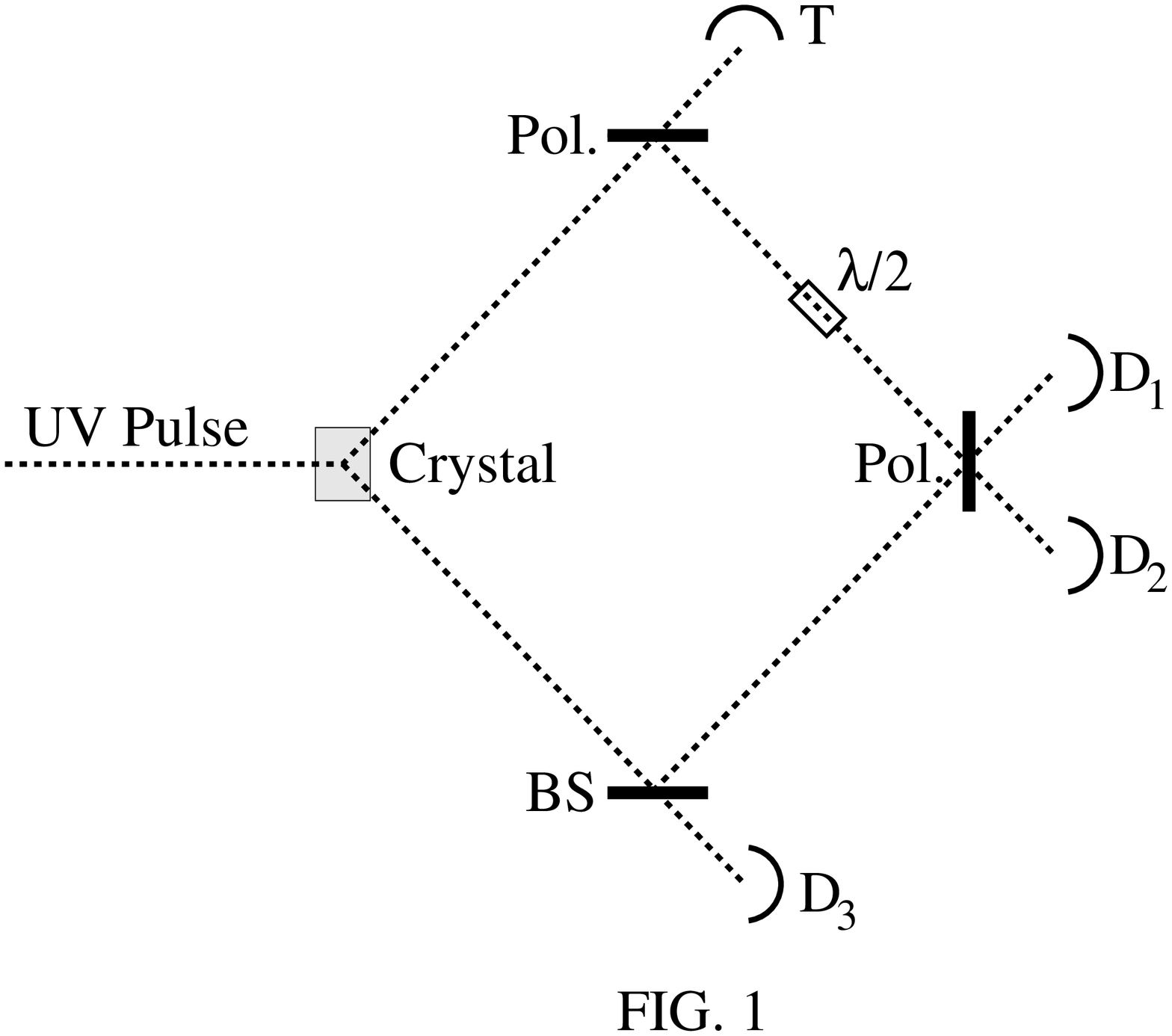}}} \par}

\caption{Scheme for the Innsbruck GHZ experiment. The GHZ correlations are obtained
when all detectors \protect\protect\( T,D_{1},D_{2},\protect \protect \) and
\protect\protect\( D3\protect \protect \) register a photon within the same
window of time. }
\end{figure}

We will assume that double pairs created have the expected GHZ correlation,
and the probability negligible of having triple pair produtions or of having
fourfold coincidence registered when no photon is generated. (Our analysis is
different from that of \.{Z}ukowski \cite{zukowski}, who considered only ideal
detectors.) Two possibilities are left: i) a pair of photons is created at the
parametric down converter; ii) two pairs of photons are created. We will denote
by \( p_{1}p_{2} \) the pair creation, and by \( p_{1}...p_{4} \) the two-pair
creation. We will assume that the probabilities add to one, i.e. \( P\left( p_{1}\ldots p_{4}\right) +P\left( p_{1}p_{2}\right) =1. \)

We start with two photons. \( p_{1}p_{2} \) can reach any of the following
combinations of detectors: \( TD_{1}, \) \( TD_{2}, \) \( TD_{3}, \) \( D_{1}D_{1}, \)
\( D_{1}D_{2}, \) \( D_{1}D_{3}, \) \( D_{2}D_{2}, \) \( D_{2}D_{3}, \)
\( D_{3}D_{3}, \) \( TT \). For an event to be counted as being a GHZ state,
all four detectors must fire (this conditionalization is equivalent to the enhancement
hypothesis). We take as our set of random variables \( \mathbf{T},\mathbf{D}_{1},\mathbf{D}_{2},\mathbf{D}_{3} \)
which take values \( 1 \) (if they fire) or \( 0 \) (if they don't fire).
We will use \( t,d_{1},d_{2},d_{3} \) (\( \overline{t},\overline{d}_{1},\overline{d}_{2},\overline{d}_{3} \))
to represent the value 1 (0). We want to compute \( P\left( td_{1}d_{2}d_{3}\mid p_{1}p_{2}\right) , \)
the probability that all detectors \( T,D_{1},D_{2},D_{3} \) fire simultaneously
given that only a pair of photons has been created at the crystal. We start
with the case when the two photons arrive at detectors \( T \) and \( D_{3}. \)
Since the efficiency of the detectors is \( d \), the probability that both
detectors detect the photons is \( d^{2}, \) the probability that only one
detects is \( 2d(1-d) \) and the probability that none of them detect is \( (1-d)^{2}. \)
Taking \( \gamma  \) into account, then the probability that all four detectors
fire is 
\[
P\left( td_{1}d_{2}d_{3}\mid p_{1}p_{2}=TD_{3}\right) =\gamma ^{2}\left( d+\gamma (1-d)\right) ^{2},\]
 where \( p_{1}p_{2}=TD_{3} \) represents the simultaneous (i.e. within a measurement
window) arrival of the photons a the trigger \( T \) and at \( D_{3}. \) Similar
computations can be carried out for \( p_{1}p_{2}=TD_{1}, \) \( TD_{2}, \)
\( D_{1}D_{3}, \) \( D_{1}D_{2}, \) \( D_{2}D_{3}. \) For \( p_{1}p_{2}=D_{i}D_{i} \)
the computation of \( P\left( td_{1}d_{2}d_{3}\mid p_{1}p_{2}=D_{i}D_{i}\right)  \)
is different. The probability that exactly one of the photons is detected at
\( D_{i} \) is \( d(1-d) \) and the probability that none of them are detected
is \( (1-d)^{2}. \) Then, it is clear that 
\[
P\left( td_{1}d_{2}d_{3}\mid p_{1}p_{2}=D_{i}D_{i}\right) =d\left( 1-d\right) \gamma ^{3}+(1-d)^{2}\gamma ^{4},\]
 and we have at once that 
\begin{eqnarray}
P\left( td_{1}d_{2}d_{3}\mid p_{1}p_{2}\right)  & = & 6\gamma ^{2}\left( d+\gamma (1-d)\right) ^{2}\nonumber \\
 &  & +4\gamma ^{3}\left( 1-d\right) \left( d+\gamma \right) .\nonumber 
\end{eqnarray}
 We note that the events involving \( P\left( td_{1}d_{2}d_{3}\mid p_{1}p_{2}\right)  \)
have no spin correlation, contrary to GHZ events.

We now turn to the case when four photons are created. The probability that
all four are detected is \( d^{4}, \) that three are detected is \( 4d^{3}(1-d), \)
that two are detected is \( 6d^{2}(1-d)^{2}, \) that one is detected is \( 4d(1-d)^{3}, \)
and that none is detected is \( (1-d)^{4}. \) If all four are detected, we
have a true GHZ-correlated state detected. However, one can again have four
detections due to dark counts. We will write \( p_{1}...p_{4}=GHZ \) to represent
having the four GHZ photons detected, and \( p_{1}...p_{4}=\overline{GHZ} \)
as having the four detections as a non-GHZ state. We can write that 
\begin{equation}
\label{21}
P\left( td_{1}d_{2}d_{3}\mid p_{1}...p_{4}=GHZ\right) =d^{4}+\gamma \left( 1-d\right) d^{3}
\end{equation}
 and 
\[
P\left( td_{1}d_{2}d_{3}\mid p_{1}...p_{4}=\overline{GHZ}\right) =3\gamma d^{3}(1-d)+6\gamma ^{2}d^{2}(1-d)^{2}+4\gamma ^{3}d(1-d)^{3}+\gamma ^{4}(1-d)^{4}.\]
 The last term in (\ref{21}) comes from the unique role of the trigger \( T, \)
that needs to detect a photon but not necessarily one that has a GHZ correlation.

How do the non-GHZ detections change the GHZ expectations?\ What is measured
in the laboratory is the conditional correlation \( E\left( \mathbf{S}_{1}\mathbf{S}_{2}\mathbf{S}_{3}\mid td_{1}d_{2}d_{3}\right)  \),
where \( \mathbf{S}_{1}, \) \( \mathbf{S}_{2} \) and \( \mathbf{S}_{3} \)
are random variables with values \( \pm 1, \) representing the spin measurement
at \( D_{1},D_{2} \) and \( D_{3} \) respectively. We can write it as 
\[
E\left( \mathbf{S}_{1}\mathbf{S}_{2}\mathbf{S}_{3}\mid td_{1}d_{2}d_{3}\right) =\frac{E\left( \mathbf{S}_{1}\mathbf{S}_{2}\mathbf{S}_{3}\mid td_{1}d_{2}d_{3}\, \, \&\, \, GHZ\right) P(GHZ)}{P(GHZ)+P(\overline{GHZ})}.\]
 since for non-GHZ states we expect a correlation zero for the term 
\[
\frac{E\left( \mathbf{S}_{1}\mathbf{S}_{2}\mathbf{S}_{3}\mid td_{1}d_{2}d_{3}\, \, \&\overline{GHZ}\right) P(\overline{GHZ})}{P(GHZ)+P(\overline{GHZ})}.\]
 Neglecting terms of higher order than \( \gamma ^{2} \), using \( \gamma \ll d \),
and \( P(p_{1}p_{2})\gg P(p_{1}...p_{4}), \) we obtain, from \( P(\overline{GHZ})=6P(p_{1}p_{2})\gamma ^{2}d^{2}+3P(p_{1}...p_{4})\gamma (1-d)d^{3} \)
and \( P\left( GHZ\right) =P(p_{1}...p_{4})\left[ d^{4}+\gamma \left( 1-d\right) d^{3}\right] , \)
that 
\begin{equation}
\label{18}
E\left( \mathbf{S}_{1}\mathbf{S}_{2}\mathbf{S}_{3}\mid td_{1}d_{2}d_{3}\right) =\frac{E(\mathbf{S}_{1}\mathbf{S}_{2}\mathbf{S}_{3}\mid td_{1}d_{2}d_{3}\&GHZ)}{\left[ 1+6\frac{P(p_{1}p_{2})}{P(p_{1}...p_{4})}\frac{\gamma ^{2}}{d^{2}}\right] }.
\end{equation}
 This value is the corrected expression for the conditional correlations if
we have detector efficiency taken into account. The product of the random variables
\( \mathbf{S}_{1}\mathbf{S}_{2}\mathbf{S}_{3} \) can take only values \( +1 \)
or \( -1. \) Then, if their expectation is \( E\left( \mathbf{S}_{1}\mathbf{S}_{2}\mathbf{S}_{3}\mid td_{1}d_{2}d_{3}\right)  \)
we have 
\[
P\left( \mathbf{S}_{1}\mathbf{S}_{2}\mathbf{S}_{3}=1\mid td_{1}d_{2}d_{3}\right) =\frac{1+E\left( \mathbf{S}_{1}\mathbf{S}_{2}\mathbf{S}_{3}\mid td_{1}d_{2}d_{3}\right) }{2}.\]
 The variance \( \sigma ^{2} \) for a random variable that assumes only \( 1 \)
or \( -1 \) values is \( 4P(1)\left( 1-P(1)\right) . \) Hence, in our case
we have as a variance 
\[
\sigma ^{2}=1-\left[ E\left( \mathbf{S}_{1}\mathbf{S}_{2}\mathbf{S}_{3}\mid td_{1}d_{2}d_{3}\right) \right] ^{2}.\]

We will estimate the values of \( \gamma  \) and \( d \) to see how much \( E\left( \mathbf{S}_{1}\mathbf{S}_{2}\mathbf{S}_{3}\mid td_{1}d_{2}d_{3}\right)  \)
would change due to experimental errors. For that purpose, we will use typical
rates of detectors \cite{EGG} for the frequency used at the Innsbruck experiment,
as well as their reported data \cite{Innsbruck}. First, modern detectors usually
have \( d\cong 0.5 \) for the wavelengths used at Innsbruck. We assume a dark-count
rate of about \( 3\times 10^{2} \) counts/s. With a time window of coincidence
measurement of \( 2\times 10^{-9} \) s, we then have that the probability of
a dark count in this window is \( \gamma \cong 6\times 10^{-7}. \) From \cite{Innsbruck}
we use that the ratio \( P(p_{1}p_{2})/P(p_{1}...p_{2}) \) is on the order
of \( 10^{10}. \) Substituting this three numerical values in (\ref{18}) we
have \( E\left( \mathbf{S}_{1}\mathbf{S}_{2}\mathbf{S}_{3}\mid td_{1}d_{2}d_{3}\right) \cong 0.9. \)
From this expression it is clear that the change in correlation imposed by the
dark-count rates is significant for the given parameters. However, it is also
clear that the value of the correlation is quite sensitive to changes in the
values of both \( \gamma  \) and \( d. \) We can now compare the values we
obtained with the ones observed by Bouwmeester et al. for GHZ and \( \overline{GHZ} \)
states \cite{Innsbruck}. In their case, they claim to have obtained a ratio
of \( 1:12 \) between \( \overline{GHZ} \) and GHZ states. In this case the
correlations are \( E\left( \mathbf{S}_{1}\mathbf{S}_{2}\mathbf{S}_{3}\mid td_{1}d_{2}d_{3}\right) \cong 0.92. \)
It is clear that a detailed analysis of the parameters would be necessary to
fit the experimental result to the predicted correlations that take the inefficiencies
into account, but at this point one can see that values close to an experimentally
measured \( 0.92 \) can be obtained with appropriate choices of the parameters
\( d \) and \( \gamma  \) (see FIG. 2).
\begin{figure}
{\par\centering \subfigure{\resizebox*{3in}{!}{\rotatebox{90}{\includegraphics{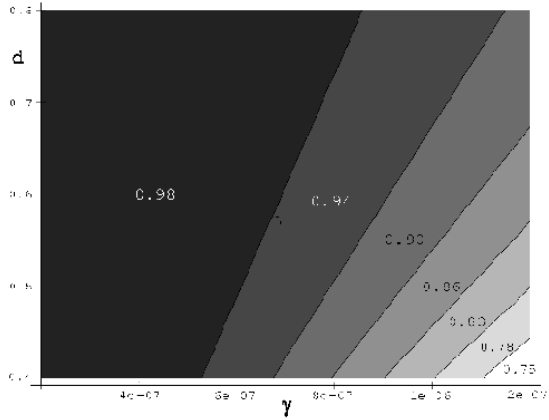}}}} \par}

\caption{Contour plot of the correlation as a function of \protect\protect\( \protect \gamma \protect \protect \)
and \protect\protect\( d.\protect \protect \) The region where the correlation
is \protect\protect\( 0.92\protect \protect \) defines a region for the parameters
\protect\protect\( \protect \gamma \protect \protect \) and \protect\protect\( d\protect \protect \)
that is compatible with the Innsbruck results. }
\end{figure}
 This expected correlation also satisfies 
\begin{equation}
\label{ineq}
E\left( \mathbf{S}_{1}\mathbf{S}_{2}\mathbf{S}_{3}\mid td_{1}d_{2}d_{3}\right) >1-\frac{1}{2}.
\end{equation}
 This result is enough to prove the nonexistence of a joint probability distribution.
We should note that the standard deviation in this case is 
\begin{equation}
\sigma \cong \sqrt{\left( 1+0.92\right) \left( 1-0.92\right) }=0.39.
\end{equation}
 As a consequence, since \( 0.92-0.39=0.53, \) the result \( 0.92 \) is bounded
away from the classical limit \( 0.5 \) by more than one standard deviation
(see FIG. 3).

We showed that the GHZ theorem can be reformulated in a probabilistic way to
include experimental inefficiencies. The set of four inequalities (\ref{inequality1})-(\ref{inequality4})
sets lower bounds for the correlations that would prove the nonexistence of
a local hidden-variable theory. Not surprisingly, detector inefficiencies and
dark-count rates can change considerably the correlations. How do these results
relate to previous ones obtained in the large literature of detector inefficiencies
in experimental tests of local hidden-variable theories. We start with Mermin's
paper \cite{Mermin2}, where an inequality for \( F \) similar to ours but
for the case of \( n \)-correlated particles is derived. Mermin does not derive
a minimum correlation for GHZ's original setup that would imply the non-existence
of a hidden-variable theory, as his main interest was to show that the quantum
mechanical results diverge exponentially from a local hidden-variable theory
if the number of entangled particles increase. Braunstein and Mann \cite{Braunstein}
take Mermin's results and estimate possible experimental errors that were not
considered here. They conclude that for a given efficiency of detectors the
noise grows slower than the strong quantum mechanical correlations. Reid and
Munru \cite{Reid} obtained an inequality similar to our first one, but there
are sets of expectations that satisfy their inequality and still do not have
a joint probability distribution. In fact, as we mentioned earlier, our complete
set of inequalities is a necessary and sufficient condition to have a joint
probability distribution.

We have used an enhancement hypothesis, namely, that we only counted events
with all four simultaneous detections, and showed that with the coincidence
constraint a joint probability did not exist in the Innsbruck experiment. Enhancement
hypotheses have to be used when detector efficiencies are low, but they may
lead to loopholes in the arguments about the nonexistence of local hidden-variable
theories. Loophole-free requirements for detector inefficiencies are based on
the analysis of \cite{Eberhard} for the Bell case and for \cite{Larsson} for
the GHZ experiment without enhancement. However, in the Innsbruck setup enhancement
is necessary, as the ratio of pair to two-pair production is of the order of
\( 10^{10} \) \cite{Innsbruck}. Until experimental methods are found to eliminate
the use of enhancement in GHZ experiments, no loophole-free results seem possible.

FIG.\ 3
\begin{figure}
{\par\centering \resizebox*{3in}{!}{\includegraphics{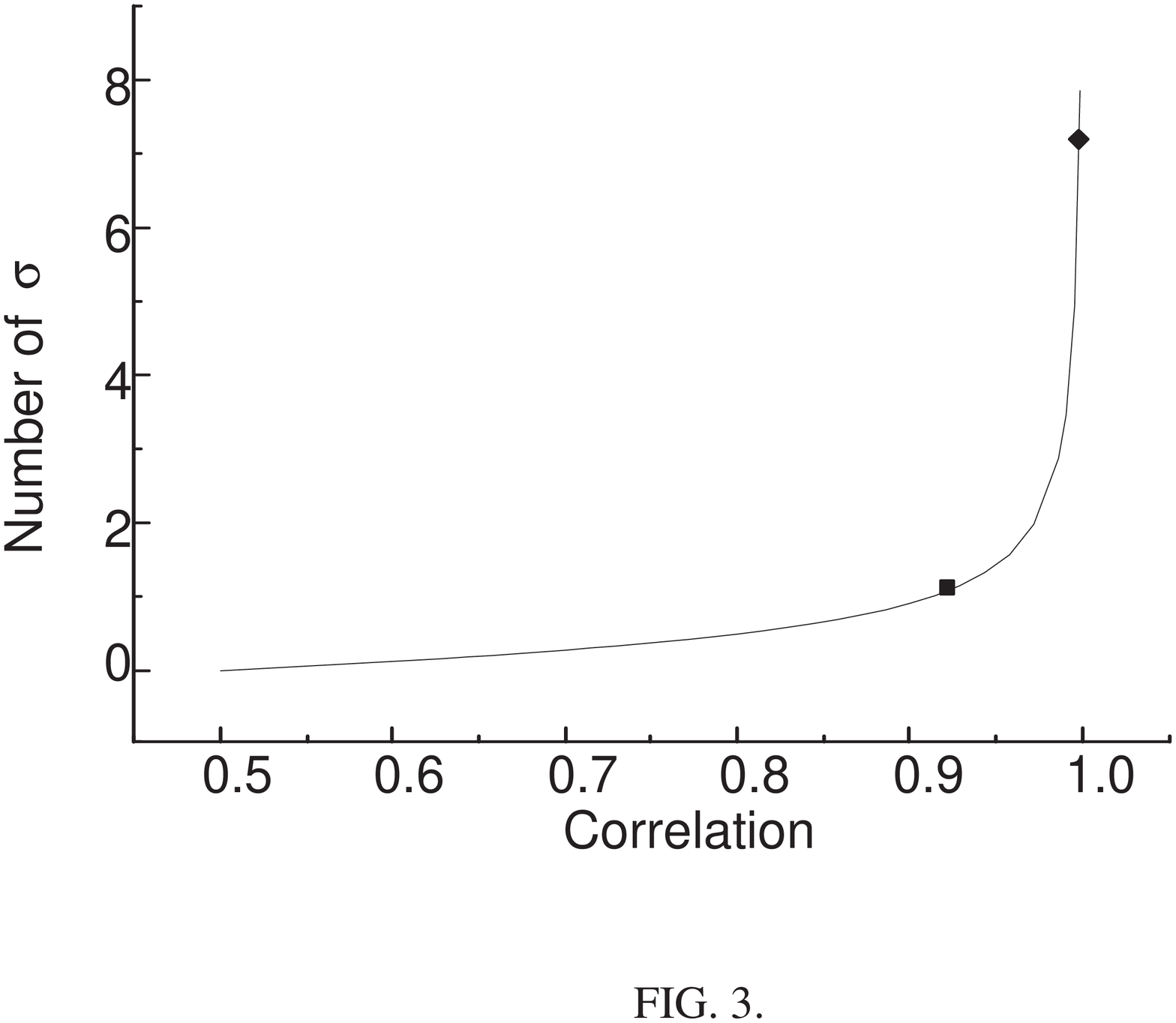}} \par}

\caption{Number of \protect\protect\( \protect \sigma \protect \protect \)'s separating
any observed correlation and the critical boundary 0.5. The square represents
the reported correlation for the Innsbruck experiment, and the diamond represents
the expected correlation if the dark count is reduced to 50 counts/s. }
\end{figure}
 shows the number of standard deviations, as computed above, by which the existence
of a joint distribution is violated. We can see that if we change the experiment
such that we reduce the dark-count rate to 50 per second, instead of the assumed
300, a large improvement in the experimental result would be expected. Detectors
with this dark-count rate and the assumed efficiency are available \cite{EGG}.
We emphasize that there are other possible experimental manipulations that would
increase the observed correlation, e.g. the ratio \( P(p_{1}p_{2})/P(p_{1}...p_{2}), \)
but we cannot enter into such details here. The point to hold in mind is that
FIG. 3 provides an analysis that can absorb any such changes or other sources
of error, not just the dark-count rate, to give a measure of reliability.

We would like to thank Prof. Sebasti\~{a}o J. N. de P\'{a}dua for comments,
as well as the anonymous referees.


\begin{thebibliography}{10}
\bibitem{GHZ}D. M. Greenberger, M. Horne, and A. Zeillinger, in \emph{Bell's Theorem, Quantum
Theory, and Conceptions of the Universe,} edited by M. Kafatos (Kluwer, Dordrecht,
1989). 
\bibitem{Mermin}N. D. Mermin, \emph{Am. J. Phys.} \textbf{58,} 731 (1990). 
\bibitem{Clauseretal}J. F. Clauser, M. A. Horne, A. Shimony, and R. A. Holt, \emph{Phys. Rev. Lett.}\textbf{\ 23,}
880 (1969). 
\bibitem{suppeszannoti}P. Suppes and M. Zanotti, \emph{Synthese} \textbf{48,} 191 (1981). 
\bibitem{Innsbruck}D. Bouwmeester, J-W. Pan, M. Daniell, H. Weingurter, and A. Zeilinger, \emph{Phys.
Rev. Lett.} \textbf{82,} 1345 (1999). 
\bibitem{zukowski}M. \.{Z}ukowski, ``Violation of Local Realism in the Innsbruck Experiment'',
quant-ph/9811013. 
\bibitem{EGG}Single photon count module specifications for EG\&G's SPCM-AG series were obtained
from EG\&G's web page at \texttt{http://www.egginc.com.}
\bibitem{Braunstein}S. Braunstein and A. Mann, \emph{Phys. Rev.} \textbf{A47,} R2427 (1993). 
\bibitem{Mermin2}N. D. Mermin, \emph{Phys. Rev. Lett.} \textbf{65}, 1838 (1990). 
\bibitem{Reid}M. D. Reid and W. J. Munro, \emph{Phys. Rev. Lett.} \textbf{69}, 997 (1992). 
\bibitem{Eberhard}P.\ H.\ Eberhard, \emph{Phys.\ Rev.\ A} \textbf{47}, R747 (1993). 
\bibitem{Larsson}J.\ Larsson, \emph{Phys.\ Rev.\ A} \textbf{57}, 3304 (1998); J.\ Larsson, \emph{Phys.\ Rev.\ A}
\textbf{57}, R3145 (1998). 
\end{thebibliography}
\end{document}